\documentclass[a4paper,12pt]{article}
\usepackage{amssymb}
\usepackage{mathrsfs}
\usepackage{bbm}
\usepackage{epsf,amsmath,amssymb}
\usepackage[dvips,usenames]{color}
\usepackage{graphicx}
\usepackage{color}
\usepackage{colortbl}
\usepackage{epsfig}

\newlength{\dinwidth}
\newlength{\dinmargin}
\def\be{\begin{equation}}
\def\ee{\end{equation}}
\def\ba{\begin{eqnarray}}
\def\ea{\end{eqnarray}}
\newcommand{\dm}{\ensuremath{\Delta M}}
\newcommand{\dg}{\ensuremath{\Delta \Gamma}}

\begin{document}

\title{\bf Implications of the Anomalies in $B_s^0-\bar{B}_s^0$ Mixing for Anomalous Tensor Couplings}
\author{Qin Chang$^{a,b}$\footnote{E-mail: changqin@htu.cn}\,, Lin Han$^{a}$\,, Ya-Dong Yang$^{a,c}$\\
{ $^a$\small Institute of Particle Physics, Huazhong Normal
University, Wuhan, Hubei  430079, P. R. China  }\\
{$^{b}$\small Department of Physics, Henan Normal University, Xinxiang, Henan 453007, P.~R. China }\\
{ $^c$\small Key Laboratory of Quark \& Lepton Physics, Ministry of Education, P.R. China}}
\date{}
\maketitle
\bigskip\bigskip
\maketitle \vspace{-1.5cm}

\begin{abstract}
Motivated by the recently observed anomalous large dimuon charge asymmetry in neutral B decays and the unexpected large CP phase in the mixing-induced CP asymmetry for $B_s\to J/\psi\phi$ decay, we study the effects of the anomalous tensor couplings to pursue possible solution. With the constraints from the obsevables $\phi^{J/\psi\phi}_s$, $a_{sl}^s$ and $\Delta M_s$, the parameter spaces are severely restricted. Numerically, we find the anomalies in $B_s^0-\bar{B}_s^0$ mixing system could be moderated simultaneously by the contributions induced by the color-singlet or color-octet tensor operators with their respective nontrivial new weak phase $\phi_{T1}\approx28.0^{\circ}$~($15.6^{\circ}$) or $\phi_{T8}\approx-62.1^{\circ}$~($-74.2^{\circ}$) and relevant strength parameters $g_{T1}\approx6.7(8.5)\times10^{-2}$ or $g_{T8}\approx1.9(2.4)\times10^{-2}$ for the CP-violating phase $\phi^{J/\psi\phi}_s=-0.77^{+0.29}_{-0.37}(-2.36^{+0.37}_{-0.29})$, respectively.
\end{abstract}

\newpage

The flavor changing neutral current~(FCNC) processes, such as $b\to s$ transitions, arise only from loop effects within the Standard Model~(SM), and therefore are very suitable for testing the SM and probing its various extensions. Among the many decay models induced by $b\to s$ transition, $B_s^0-\bar{B}_s^0$ oscillation is of great importance, which is governed by a Schr\"{o}dinger equation
\begin{equation}
i \frac{d}{dt}
\left(
\begin{array}{c}
|B_s(t)\rangle \\ |\bar{B}_s (t)\rangle
\end{array}
\right)
=
\left( M^s - \frac{i}{2} \Gamma^s \right)
\left(
\begin{array}{c}
|B_s(t)\rangle \\ |\bar{B}_s (t)\rangle
\end{array}
\right)\label{sch}
\end{equation}
with the mass matrix $M^s$ and the decay matrix $\Gamma^s$. The mass and the width differences between the light and the heavy $B_s$ mass eigenstates~($B_L$ and $B_H$) are
\begin{eqnarray}
\dm_s &=& M^s_H -M^s_L \; = \; 2\, |M_{12}^s|,
\qquad \dg_s \; =\; \Gamma^s_L-\Gamma^s_H \; =\;
        2\, |\Gamma_{12}^s| \cos \phi_s, \label{dmdg}
\end{eqnarray}
with the phase $\phi_s=\arg(-M_{12}^s/\Gamma_{12}^s)$. Another independent observable is the like-sign dimuon charge asymmetry for the semileptonic decay of $B_{s,d}^0$ mesons, which is defined by
\begin{eqnarray}
A_{sl}^{b}\equiv\frac{N^{++}_b-N^{--}_b}{N^{++}_b+N^{--}_b}=\beta_da^d_{\rm sl}+\beta_sa^s_{\rm sl}\,,\quad{\rm with }\quad
a^q_{\rm sl}={\rm Im} \frac{\Gamma_{12}^q}{M_{12}^q}=\frac{|\Gamma_{12}^q|}{|M_{12}^q|} \sin \phi_q\,, \label{deassl}
\end{eqnarray}
where $N^{++}_b$ and $N^{--}_b$ represent the numbers of events containing two $B_{s,d}^0$ mesons decaying semileptonically into two positive or two negative muons, respectively.

Within the SM, the theoretical evaluations for these quantities have been fully studied~\cite{Buchalla:1996vs,Lenz,beneke}. The recent updated SM predictions for the mass and the width differences are $\dm_s=(19.30\pm6.68){\rm ps}^{-1}$ and $\dg_s=(0.096\pm0.039){\rm ps}^{-1}$~\cite{Lenz},  which agree well with the CDF Collaboration result for $\dm_s$~\cite{CDFdms} and Heavy Flavor Averaging Group~(HFAG) averaged data for $\dg_s$~\cite{HFAG}
\ba
\dm_s&=&17.77\pm0.10\pm0.07\,{\rm ps}^{-1}\,,\quad {\rm CDF}\\
\label{dgsHFAG}
\dg_s&=&0.154^{+0.054}_{-0.070}\,{\rm or}\,-0.154_{-0.054}^{+0.070}\,{\rm ps}^{-1}\,,\quad {\rm HFAG}
\ea
respectively. While, some recent measurements of CP violating observations in $B_s$ mixing system present some anomalies, which may be the hints for new physics~(NP) and motivate the search for new source of CP violation.

One of the hints for the existence of new CP violating source in $B_s^0-\bar{B}_s^0$ system is an unexpected large CP phase in the mixing-induced CP asymmetry for $B_s\to J/\psi\phi$ measured by CDF~\cite{CDFbs,CDFnew} and D0~\cite{D0bs,D0new} Collaborations. Averaging the data from CDF and D0, two possible solutions~( named S1 and S2 for convenience) for the CP phase $\phi_s^{J/\psi\phi}$~\footnote{Theoretically, the phase $\phi_s^{J/\psi\phi}$ is different from $\phi_s$ appeared in Eq.~(\ref{deassl}) (see Ref.~\cite{Lenz} for detail).} are given by  HFAG ~\cite{HFAG}

\be\label{phijp}
\phi_s^{J/\psi\phi}=\left\{\begin{array}{l}
-0.77^{+0.29}_{-0.37}\quad({\rm S1})\,,\\
-2.36^{+0.37}_{-0.29}\quad({\rm S2})\,.\end{array}\right.
\ee
However, within the SM, the CP violating phase is predicted to be small
\be
\phi_s^{J/\psi\phi}(SM)=2{\rm arg}[-V_{tb}V_{ts}^{\ast}/V_{cb}V_{cs}^{\ast}]\approx-0.040\,,
\ee
which deviates from experimental data by more than $2.5\sigma$.

Another hint for new source of CP violating is the recently observed anomalously large CP-violating like-sign dimuon charge asymmetry in semileptonic B hadron decays reported by D0 Collaboration~\cite{D0}. Using the data corresponding to $6.1 fb^{-1}$of integrated luminosity, D0 Collaboration reports the result~\cite{D0}
\be
A_{sl}^b=(9.57\pm2.51(\rm stat)\pm1.46(sys))\times10^{-3}\,,
\ee
which differs by $3.2\sigma$ from the SM prediction $(-2.3^{+0.5}_{-0.6})\times10^{-4}$~\cite{Lenz}.
With the known values for mixing parameters of $B_{d,s}$ system~\cite{PDG10}, D0 Collaboration obtains~\cite{D0}
\be\label{Aslb}
A_{sl}^b=(0.506\pm0.043)a_{sl}^d+(0.494\pm0.043)a_{sl}^s\,.
\ee
Using the known experiment value $a_{sl}^d=(-4.7\pm4.6)\times10^{-3}$~\cite{HFAG}, Eq.~(\ref{Aslb}) leads to $a_{sl}^s=(-14.6\pm7.5)\times10^{-3}$~\cite{D0}. Combining CDF Collaboration measurement of $A_{sl}^b$~\cite{CDFAslb} and D0 Collaboration direct measurement of $a_{sl}^s$~\cite{D0asls}, one can get the average value
\be\label{asls}
a_{sl}^s=(-12.7\pm5.0)\times10^{-3}\,,
\ee
which deviates from the SM prediction $(2.06\pm0.57)\times10^{-5}$\cite{Lenz} by $2.5\sigma$. If such a large $a_{sl}^s$ is confirmed, it would imply the existence of a significant new source of the CP violation beyond the SM in the $B_s^0-\bar{B}_s^0$ mixing system.

In many extension of the SM, various new couplings could be generated and represented by new four-quark operators in effective theory. One kind of them is the anomalous tensor operators, which are helpful to resolve the abnormally large transverse polarizations observed in $B\to \phi K^\ast$ decay, as well as the large ${\cal B}(B\to\eta K^\ast)$~\cite{ChangT,YYtensor,KC} and have attracted  many attentions recently~\cite{tensor}. In this Letter, motivated by the aforementioned anomalies in $B_s^0-\bar{B}_s^0$ mixing system, we shall pursue possible solutions through a set of anomalous FCNC tensor operators in a model independent way.

The full set of the tensor operators responsible for the $B_s^0-\bar{B}_s^0$ mixing belonging to the TLL, TRR and TLR sectors read
\ba
\label{TO1}
{\rm TLL:}\quad O_{T1}^{LL}=\bar{s}_i\sigma_{\mu\nu}(1-\gamma_5)b_i\bar{s}_j\sigma^{\mu\nu}(1-\gamma_5)b_j\,,\quad O_{T8}^{LL}=\bar{s}_i\sigma_{\mu\nu}(1-\gamma_5)b_j\bar{s}_j\sigma^{\mu\nu}(1-\gamma_5)b_i\,;\\
\label{TO2}
{\rm TRR:}\quad O_{T1}^{RR}=\bar{s}_i\sigma_{\mu\nu}(1+\gamma_5)b_i\bar{s}_j\sigma^{\mu\nu}(1+\gamma_5)b_j\,,\quad O_{T8}^{RR}=\bar{s}_i\sigma_{\mu\nu}(1+\gamma_5)b_j\bar{s}_j\sigma^{\mu\nu}(1+\gamma_5)b_i\,;\\
\label{TO3}
{\rm TLR:}\quad O_{T1}^{LR}=\bar{s}_i\sigma_{\mu\nu}(1-\gamma_5)b_i\bar{s}_j\sigma^{\mu\nu}(1+\gamma_5)b_j\,,\quad O_{T8}^{LR}=\bar{s}_i\sigma_{\mu\nu}(1-\gamma_5)b_j\bar{s}_j\sigma^{\mu\nu}(1+\gamma_5)b_i\,,
\ea
where $i$ and $j$ are color indices. For the TLL operators, because the renormalization group~(RG) evolution and the parametrization of the matrix elements are the same as TRR sector, its contributions could be easily obtained by replacing the corresponding NP parameters; As for the TLR operators, the hadronic matrix elements relevant to the two operators $O_{T1,8}^{LR}$ are zero in vacuum insertion approach. So, in this Letter, we shall pay our attention only to the TRR operators for simplicity.

To begin with, at electro-weak~(EW) scale, the four fermion interactions responsible for $B_s^0-\bar{B}_s^0$ mixing induced by the tensor operators $O_{T1}^{RR}$ and $O_{T8}^{RR}$ could be described by the effective Lagrangian
\begin{equation}\label{Leff1}
 {\cal L}_{eff}^{T}=-G_F^2(g_{T1}^2 O_{T1}^{RR}+g_{T8}^2 O_{T8}^{RR})+{\rm h.c.}\,,
\end{equation}
where the new effective chiral $b-s$ FCNC tensor couplings $g_{T1}$ and $g_{T8}$ are generally complex and could be written as
\ba\label{oper}
g_{T1}\equiv |g_{T1}|e^{i\phi_{T1}}\,,\quad g_{T8}\equiv |g_{T8}|e^{i\phi_{T8}}\,.
\ea
Using the color-singlet operators basis introduced in Ref.~\cite{BurasDF2},
the effective Lagrangian Eq.~(\ref{Leff1}) could be rewritten as
\begin{equation}\label{Leff2}
 {\cal L}_{eff}^{T}=-G_F^2\big[(g_{T1}^2+\frac{1}{2}g_{T8}^2) O_{T1}^{RR}-6g_{T8}^2O_{S1}^{RR}\big]+{\rm h.c.}\,,
\end{equation}
through the Fierz identity
\be
O_{T8}^{RR}=-6O_{S1}^{RR}+\frac{1}{2}O_{T1}^{RR}\,,
\ee
where $O_{S1}^{RR}=\bar{s}_i(1+\gamma_5)b_i\bar{s}_j(1+\gamma_5)b_j$.
Then, including both the SM and the tensor operators contributions, the full expression of the effective Hamiltonian for $\Delta B=\Delta S=2$ transition at $m_b$ scale could be written as
\begin{equation}\label{Heff}
 {\cal H}_{eff}^{\prime}(\mu_b)=\frac{G_F^2}{16\pi^2}m_W^2(V_{tb}V^{\ast}_{ts})^2\big[C_V^{LL}(\mu_b)O_V^{LL}+C_{T1}^{RR}(\mu_b)O_{T1}^{RR}+C_{S1}^{RR}(\mu_b)O_{S1}^{RR}\big]+{\rm h.c.}\,,
\end{equation}
where $O_V^{LL}=\bar{s}_i\gamma_{\mu}(1-\gamma_s)b_i\bar{s}_j\gamma^{\mu}(1-\gamma_s)b_j$, $C_{V}^{LL}(\mu_b)$ and $C_{T1,S1}^{RR}(\mu_b)$ are the Wilson coefficients at the scale $\mu_b=m_b$. In Eq.~(\ref{Heff}), the first term corresponds to the SM part, and the last two terms are the contributions induced by the tensor couplings. At the EW scale $\mu_W=M_W$, the Wilson coefficients read
\ba
C_V^{LL}(\mu_{W})&=&S_0(x_t)+\frac{\alpha_s(\mu_{W})}{4\pi}\big[S_1(x_t)+F(\mu_W)S_0(x_t)+B_t S_0(x_t)\big]\,,~\cite{Buchalla:1996vs}\\
C_{T1}^{RR}(\mu_{W})&=&\frac{16\pi^2}{M_W^2}\,\frac{|g_{T1}|^2e^{i2\phi_{T1}}+\frac{1}{2}|g_{T8}|^2e^{i2\phi_{T8}}}{(V_{tb}V^{\ast}_{ts})^2}\,,\\
C_{S1}^{RR}(\mu_{W})&=&\frac{16\pi^2}{M_W^2}\,\frac{-6|g_{T8}|^2e^{i2\phi_{T8}}}{(V_{tb}V^{\ast}_{ts})^2}\,.
\ea
For convenience, in the following we will absorb the factor $\frac{4\pi}{M_W}$ into the effective couplings parameters $g_{T1}$ and $g_{T8}$. The Renormalization Group~(RG) evolution of these Wilson coefficients from the $\mu_W$ scale down to $\mu_b$ scale have been fully developed in Ref.~\cite{BurasDF2}.
With the NLO $\eta$ factors given by Ref.~\cite{BurasDF2}, the explicit expressions for the Wilson coefficients at $\mu_b$ scale are~\cite{BurasDF2}
\ba
C_V^{LL}(\mu_{b})&=&[\eta(\mu_b)]_{VLL}C_V^{LL}(\mu_{W})\,,\nonumber\\
\left(\begin{array}{c}
C_{S1}^{RR}(\mu_{b})   \\
C_{T1}^{RR}(\mu_{b})
\end{array}\right)
&=&
\left(\begin{array}{cc}
\left[\eta_{11} (\mu_b)\right]_{\rm SRR} &
\left[\eta_{12} (\mu_b)\right]_{\rm SRR} \\
\left[\eta_{21} (\mu_b)\right]_{\rm SRR} &
\left[\eta_{22} (\mu_b)\right]_{\rm SRR}
\end{array}\right)
\left(\begin{array}{c}
C_{S1}^{RR}(\mu_{W})  \\
C_{T1}^{RR}(\mu_{W})
\end{array}\right)\,.
\ea
The hadronic matrix elements corresponding to the operators in Eq.~(\ref{Heff}) could be parameterized as
\ba
\langle O_V^{LL}(\mu_b)\rangle&=&\frac{8}{3}m_{B_s}^2f_{B_s}^2B_1(\mu_b)\,,\\
\langle O_{T1}^{RR}(\mu_b)\rangle&=&-\frac{4}{3}\Big(\frac{m_{B_s}}{m_b(\mu_b)+m_s(\mu_b)}\Big)^2m_{B_s}^2f_{B_s}^2\big(2B_3(\mu_b)-5B_2(\mu_b)\big)\,,\\
\langle O_{S1}^{RR}(\mu_b)\rangle&=&-\frac{5}{3}\Big(\frac{m_{B_s}}{m_b(\mu_b)+m_s(\mu_b)}\Big)^2m_{B_s}^2f_{B_s}^2B_2(\mu_b)\,.
\ea

Finally, in terms of the effective Hamiltonian given in Eq.~(\ref{Heff}), the off-diagonal $M_{12}^s$ in the $B_s$ mass matrix is given by
\ba
\label{M12s}
2m_{B_s}M_{12}^{s}&=&\langle B_s^0|{\cal H}_{eff}^{\prime}|\bar{B}_s^0\rangle\,\nonumber\\
&=&\frac{G_F^2}{16\pi^2}M_W^2(V_{tb}V_{ts}^{\ast})^2\big[C_V^{LL}(\mu_{b})\langle O_V^{LL}(\mu_b)\rangle\,\nonumber\\
&&\,+\,C_{T1}^{RR}(\mu_{b})\langle O_{T1}^{RR}(\mu_b)\rangle\,+\,C_{S1}^{RR}(\mu_{b})\langle O_{S1}^{RR}(\mu_b)\rangle \big]\,,
\ea
where the first term is the SM contribution for $M_{12}^{s}$ and could be rewritten as~\cite{Buchalla:1996vs}
\ba
\label{M12SM}
2m_{B_s}M_{12}^{s}(SM)=\frac{G_F^2}{6\pi^2}M_W^2(V_{tb}V_{ts}^{\ast})^2(\hat{B}_{B_s}f_{B_s}^2)
m_{B_s}^2 \eta_{B}S_{0}(x_t)\,.
\ea

In the SM, the off-diagonal element of the decay matrix $\Gamma_{12}^s$ have been fully evaluated in Refs.~\cite{Lenz,beneke}
\begin{eqnarray}
\Gamma_{12}^s(SM) &=& - [\, \lambda_c^2 \, \Gamma_{12}^{cc} \; + \;
        2 \, \lambda_c\,  \lambda_u \, \Gamma_{12}^{uc}\; + \;
        \lambda_u^{2} \, \Gamma_{12}^{uu} \,  ] \label{ga12} \nonumber\\
      &=&
    - [\, \lambda_t^2 \, \Gamma_{12}^{cc} \; + \;
        2 \, \lambda_t\,  \lambda_u \,
        ( \Gamma_{12}^{cc} - \Gamma_{12}^{uc} ) \; + \;
        \lambda_u^{2} \,
        ( \Gamma_{12}^{cc} - 2 \Gamma_{12}^{uc} + \Gamma_{12}^{uu})
       ] \label{ga12t}
\end{eqnarray}
with the CKM factors $\lambda_i=V_{is}^* V_{ib}$ for $i=u,c,t$. The explicit expressions for $\Gamma_{12}^{cc,uu,uc}$ could be found in Refs.~\cite{Lenz,beneke}.
It is important to note that the NP operators Eq.~(\ref{TO1}) considered in this Letter can significantly affect $M_{12}^{s}$ but not $\Gamma_{12}^s$, which is dominated by the CKM favored $b\to c\bar{c}s$ tree-level decays within the SM. Hence $\Gamma_{12}^s=\Gamma_{12}^s(SM)$ holds as a good approximation.

\begin{table}[t]
 \begin{center}
 \caption{The values of the theoretical input parameters.}
 \label{input}
 \vspace{0.2cm}
 \doublerulesep 0.7pt \tabcolsep 0.1in
 \begin{tabular}{lccccccccccc} \hline \hline
 $\alpha_s(M_z)=0.1184$\,, $G_F=1.16637\times10^{-5}\,{\rm GeV}^{-2}$\,,  $m_W=80.399\,{\rm GeV}$\,,$m_{B_s}=5.366\,{\rm GeV}$\,,  \\
 $\bar{m}_s(2{\rm GeV})=0.101_{-0.021}^{+0.029}\,{\rm GeV}$\,, $\bar{m}_c(\bar{m}_c)=1.27^{+0.07}_{-0.09}\,{\rm GeV}$\,,$\bar{m}_b(\bar{m}_b)=4.20^{+0.17}_{-0.07}\,{\rm GeV}$\,,\\
 $m_b^{pole}=4.79^{+0.19}_{-0.08}$\,, $m_t^{pole}=172.4\pm1.2\,{\rm GeV}$\,,$\bar{m}_t(\bar{m}_t)=164.8\pm1.2\,{\rm GeV}$\,,&~\cite{PDG10,PMass}\\  \hline
 $A=0.804\pm0.010$\,, $\lambda=0.22535\pm0.00065$\,, $\bar{\rho}=0.135\pm0.040$\,, $\bar{\eta}=0.374\pm0.026$\,,&~\cite{UTfit}\\\hline
 $f_{B_{s}}=(0.231\pm0.015)\,{\rm GeV}$\,,\\
 $B_1(\mu_b)=0.86\pm0.02^{+0.05}_{-0.04}$\,, $B_2(\mu_b)=0.83\pm0.02\pm0.04$\,, $B_3(\mu_b)=1.03\pm0.04\pm0.09$\,.&~\cite{DecayCon,BagPara} \\
  \hline \hline
 \end{tabular}
 \end{center}
 \end{table}

\begin{table}[t]
 \begin{center}
 \caption{Numerical results for $\Delta M_s[{\rm ps^{-1}}]$, $\Delta \Gamma_s[\times 10^{-2}\,{\rm ps^{-1}}]$, $\phi^{J/\psi\phi}_s$ and $a_{sl}^s(\times10^{-3})$  within the SM and the anomalous tensor couplings.   }
 \label{tabpred}
 \vspace{0.5cm}
 \footnotesize
 \doublerulesep 0.7pt \tabcolsep 0.05in
 \begin{tabular}{lccccccccccc} \hline \hline
                       &Exp.                                           & SM     &\multicolumn{2}{c}{Case A}       &\multicolumn{2}{c}{Case B}\\
                       &                                               &                 & S1            &S2             & S1            &S2\\\hline
 $\Delta M_s$          &$17.77\pm0.12$                                 &$18.25\pm3.69$   &$17.77\pm0.20$ &$17.77\pm0.20$ &$17.77\pm0.20$ &$17.77\pm0.20$\\
 $\Delta \Gamma_s$     &$15.4^{+5.4}_{-7.0}$ (or $-15.4^{+7.0}_{-5.4}$)&$9.5\pm4.0$      &$5.9\pm4.7$    &$-4.8\pm4.4$   &$5.5\pm4.3$    &$-4.8\pm4.5$\\
 $\phi^{J/\psi\phi}_s$ &$-0.77^{+0.29}_{-0.37}\cup-2.36^{+0.37}_{-0.29}$&$-0.040\pm0.004$&$-1.08\pm0.39$ &$-2.07\pm0.42$ &$-1.06\pm0.36$ &$-2.05\pm0.40$\\
 $a_{sl}^s$            &$-12.7\pm5$                                    &$0.026\pm0.009$  &$-5.7\pm1.4$   &$-5.7\pm1.4$   &$-5.4\pm1.1$   &$-5.6\pm1.3$\\
 \hline \hline
 \end{tabular}
 \end{center}
 \end{table}

With the relevant theoretical formulae given above and the values of the input parameters summarized in Table~\ref{input}, we now proceed to present our numerical analyses and discussions. Including all of the theoretical uncertainties, our results of the SM predictions are listed in the third column of Table~\ref{tabpred}. One may easily find that our results $\Delta M_s=18.25\pm3.69\,{\rm ps^{-1}}$ and $\Delta \Gamma_s=(9.5\pm4.0)\times 10^{-2}{\rm ps^{-1}}$ agree well with the experimental data $17.77\pm0.12{\rm ps^{-1}}$ and $(15.4^{+5.4}_{-7.0})\times 10^{-2}{\rm ps^{-1}}$, respectively. However, the predictions $\phi^{J/\psi\phi}_s=-0.040\pm0.004$ and $a_{sl}^s=(2.6\pm0.9)\times10^{-5}$, which agree with the former SM results~(for example, the ones in Ref.~\cite{Lenz}), significantly deviate from experimental results $-0.77^{+0.29}_{-0.37}\cup-2.36^{+0.37}_{-0.29}$ and $(-12.7\pm5)\times10^{-3}$, respectively. In the following, we shall pursue possible solutions with the anomalous tensor couplings.

For the convenience of analysis, with the central values of the theoretical input parameters listed in Table.~\ref{input}, the amplitude of $B_s^0-\bar{B}_s^0$ mixing could be written as
\ba
\label{Afull}
\langle B_s|{\cal H}_{eff}^{\prime}(\mu_b)|\bar{B}_s\rangle&\equiv&{\cal A}_V^{LL}(SM)+{\cal A}_{T1}^{RR}+{\cal A}_{T8}^{RR}\,,\\
\label{AS}
{\cal A}_V^{LL}(SM)\times10^{11}&=&(6.26-i0.25)\,,\\
\label{AT1}
{\cal A}_{T1}^{RR}\times10^{11}&=&-0.148\times|g_{T1}\times10^{2}|^2e^{i2\phi_{T1}}\,,\\
\label{AT8}
{\cal A}_{T8}^{RR}\times10^{11}&=&1.83\times|g_{T8}\times10^{2}|^2e^{i2\phi_{T8}}\,.
\ea
It is found that the contributions of the tensor operators ${\cal A}_{T1}^{RR}$ and ${\cal A}_{T8}^{RR}$ with $|g_{T1,T8}|\sim{\cal O}(10^{-2})$ could be comparable with the SM contribution ${\cal A}_V^{LL}(SM)$ and may resolve the anomalies in the $B_s^0-\bar{B}_s^0$ mixing.

\begin{figure}[t]
\begin{center}
\epsfxsize=15cm \centerline{\epsffile{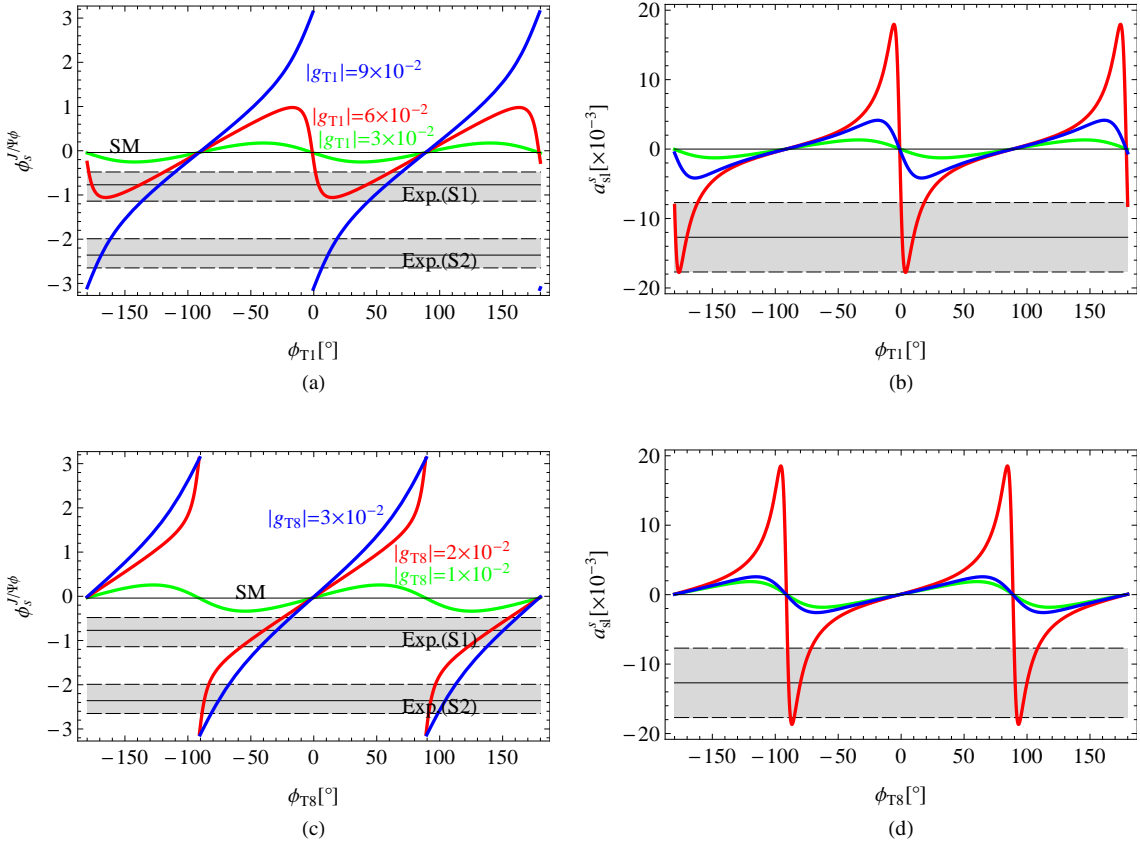}}
\centerline{\parbox{16cm}{\caption{\label{obsnp}\small The dependance of $\phi^{J/\psi\phi}_s$ and $a_{sl}^s$ on the new phases $\phi_{T1,T8}$ with different $|g_{T1,T8}|$ values. For each plot, the irrelevant NP parameters are set to zero. The dashed lines correspond to the error bars~($1\sigma$). The legends for figures (b) and (d) are the same as the ones labeled in figures (a) and (c) respectively. }}}
\end{center}
\end{figure}

With the central values of the theoretical input parameters, the dependance of the observables $\phi^{J/\psi\phi}_s$ and $a_{sl}^s$ on the new phases $\phi_{T1,T8}$ with different $|g_{T1,T8}|$ values are shown in Fig.~\ref{obsnp}. From Fig.~\ref{obsnp}~(a), we find that $\phi^{J/\psi\phi}_s$ could be brought to the experimental level of S1 with $\phi_{T1}\sim20^{\circ}$ and $|g_{T1}|\sim6\times10^{-2}$. A larger $|g_{T1}|$ is demanded to enhance $\phi^{J/\psi\phi}_s$ to the experimental result of S2 than that of S1. Meanwhile, with the same values of $|g_{T1}|$ and $\phi_{T1}$, from Fig.~\ref{obsnp}~(b), it is interesting to note that $a_{sl}^s$ also could agree with the experimental result by the contribution of ${\cal A}_{T1}^{RR}$. For the effects of ${\cal A}_{T8}^{RR}$, as shown in Figs.~\ref{obsnp}~(c) and (d), a similar situation could be found with a smaller $|g_{T8}|\sim2\times10^{-2}$ and a negative phase $\phi_{T8}\sim-65^{\circ}$. Therefore, the large discrepancy between the SM prediction and experimental measurement of both $\phi^{J/\psi\phi}_s$ and $a_{sl}^s$ could be moderated simultaneously by the contributions of $O_{T1}^{RR}$ and/or $O_{T8}^{RR}$ operators.

From Figs.\ref{obsnp}~(b) and (d), one may find that large $|g_{T1}|$~($\sim9\times10^{-2}$, for example) or $|g_{T8}|$~($\sim3\times10^{-2}$, for example) would lead to much larger contributions to $M_{12}^s$, however, they are disfavored by the anomaly $a_{sl}^s$, which also can be found from Eq.~(\ref{deassl}). Furthermore, such large $|g_{T1}|$ and/or $|g_{T8}|$ are also very fragile under the constraint from the well measured $\Delta M_s$, which SM prediction agrees well with the experimental data. Therefore, combining the constraints from $\phi^{J/\psi\phi}_s$, $a_{sl}^s$ and $\Delta M_s$, the NP parameter space is restricted very much. So, in order to evaluate the possible strength of the NP effects, we shall perform a detailed numerical analysis in the following.

\begin{figure}[t]
\begin{center}
\epsfxsize=15cm \centerline{\epsffile{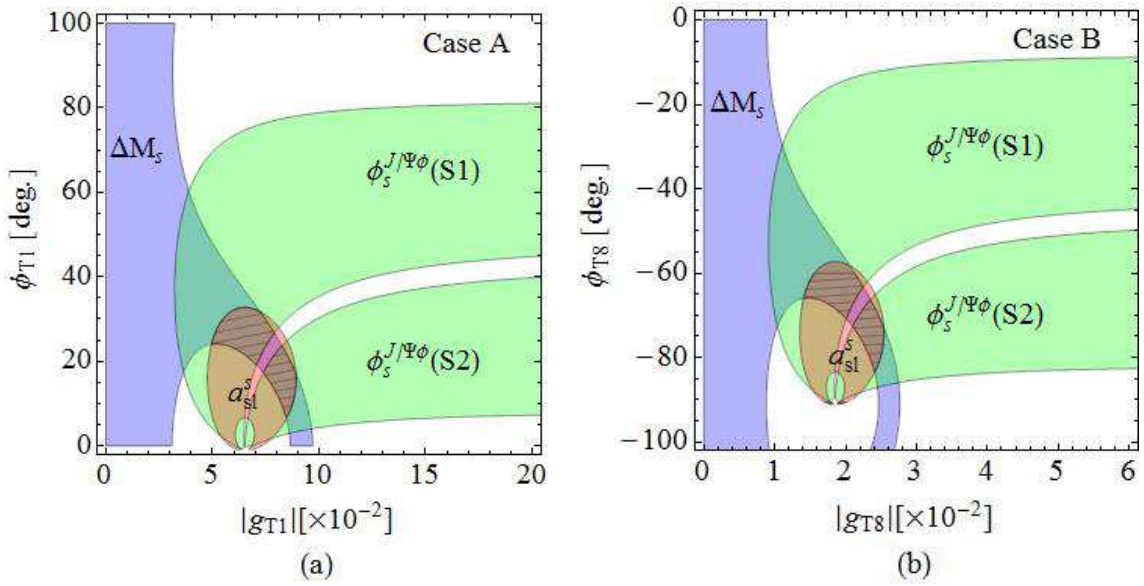}}
\centerline{\parbox{16cm}{\caption{\label{NPrange}\small The allowed regions of the tensor couplings parameters $|g_{T1,T8}|$ and $\phi_{T1,T8}$ in case~A and B under the constraints from $\phi^{J/\psi\phi}_s$~(green region), $a_{sl}^s$~(pink region), $\Delta M_s$~(blue region) and their combination~(meshed region).}}}
\end{center}
\end{figure}
\begin{table}[t]
 \begin{center}
 \caption{Fitting results for the tensor couplings parameters $|g_{T1,T8}|[\times10^{-2}]$ and
 $\phi_{T1,T8}[^{\circ}]$. }
 \label{NPpara}
 \vspace{0.5cm}
 \doublerulesep 0.7pt \tabcolsep 0.1in
 \begin{tabular}{lccccccccccc} \hline \hline
 NP pars.&\multicolumn{2}{c}{Case A} &\multicolumn{2}{c}{Case B} \\
              &S1           &S2            &S1            &S2\\\hline
 $|g_{T1}|$   &$6.7\pm1.8$  &$8.5\pm1.4$   &---           &---             \\
 $\phi_{T1}$  &$28.0\pm7.1$ &$15.6\pm7.3$  &---           &---             \\\hline
 $|g_{T8}|$   &---          &---           &$1.9\pm0.4$   &$2.4\pm0.3$         \\
 $\phi_{T8}$  &---          &---           &$-62.1\pm6.9$ &$-74.2\pm7.1$         \\
  \hline \hline
 \end{tabular}
 \end{center}
 \end{table}

Due to the cancelation between the contributions of ${\cal A}_{T1}^{RR}$ and ${\cal A}_{T8}^{RR}$, which can be found in Eqs.~(\ref{AT1}) and (\ref{AT8}), tensor couplings $g_{T1}$ and $g_{T8}$ are hardly to be well bounded without any further simplification. In order to evaluate their respective effects, our following calculations and discussions are divided into two simplified color scenarios named case~A and case~B, which correspond to with only color-singlet or -octet tensor operator considered, respectively. In each case, our fitting for the tensor couplings is performed with the experimental data on $\phi^{J/\psi\phi}_s$, $a_{sl}^s$ and $\Delta M_s$ within $1.68\sigma$~(about $90\%$ C.L.) as constraints and the theoretical inputs allowed within their respective uncertainties listed in Table.~\ref{input}. As for the observable $\dg_s$, due to its weak constraint on the tensor couplings with its large experimental error bar, we leave it as our prediction, which could be tested by the refined measurement in the forthcoming years. The allowed regions of the tensor couplings parameters $|g_{T1,T8}|$ and $\phi_{T1,T8}$ in the two cases under the constraints from $\phi^{J/\psi\phi}_s$, $a_{sl}^s$, $\Delta M_s$ and their combination are shown in Fig.~\ref{NPrange}, and the corresponding numerical results are listed in Table~\ref{NPpara}, where the two solutions S1 and S2 correspond to the two experimental results of $\phi^{J/\psi\phi}_s$ given in Eq.~(\ref{phijp}). Our theoretical results for the observables are summarized in Table~\ref{tabpred}.

\subsubsection*{Case~A: only color-singlet operator}
In order to evaluate the effects of color-singlet tensor operator, we neglect the NP contributions involving $g_{T8}$. Under the constraints from $\phi^{J/\psi\phi}_s$, $a_{sl}^s$ and $\Delta M_s$, Fig.~\ref{NPrange}~(a) shows the allowed spaces of the tensor coupling parameters. We find all of these constraint from $\phi^{J/\psi\phi}_s$, $a_{sl}^s$ and $\Delta M_s$ are improtant to get the restricted NP parameter space. Numerically, the tensor coupling parameters are seriously bounded to $|g_{T1}|=(6.7\pm1.8)\times10^{-2}$ ($(8.5\pm1.4)\times10^{-2}$) and $\phi_{T1}=28.0^{\circ}\pm7.1^{\circ}$ ($15.6^{\circ}\pm7.3^{\circ}$) in S1~(S2), which are nontrivial to moderate the discrepancies of $\phi^{J/\psi\phi}_s$ and $a_{sl}^s$ between the SM predictions and the experimental measurements.

Corresponding to the values of  $|g_{T1}|$ and $\phi_{T1}$ in S1 and S2, our theoretical results of the observables are listed in the fourth and the fifth columns of  Table~\ref{tabpred}. We find our result of $\phi^{J/\psi\phi}_s$ agrees well with the measurement within $1\sigma$. Meanwhile, $a_{sl}^s$ is enhanced to the experimental level by the contributions from ${\cal A}_{T1}^{RR}$, but it also deviates from the experimental data by about $1.1\sigma$, which is easily understood through the following analysis.

From Eqs.~(\ref{dmdg}) and~(\ref{deassl}), one can obtain
\begin{eqnarray}
a^s_{\rm sl}=\frac{|\Gamma_{12}^s|}{\dm_s} 2 \sin \phi_s\,, \label{assla}
\end{eqnarray}
for which, we can take ${\dm_s}\simeq17.77$ due to the well measurement and agreement with SM prediction, and $|\Gamma_{12}^s|=|\Gamma_{12}^s|_{SM}\simeq0.051$ due to the assumption of the negligible NP contributions to $\Gamma_{12}^s$. With the central value of the experimental result for $a^s_{\rm sl}$ as input, using Eq.~(\ref{assla}), one can get
\be
-12.7\times10^{-3}\simeq\frac{0.051}{17.77}\,2\, \sin\phi_s\Rightarrow  \sin\phi_s\simeq-2\,,
\ee
which is obviously out of the allowed $\sin\phi_s$ values. So, $a^s_{\rm sl}$ is hardly to be enhanced to around $-12.7\times10^{-3}$ by NP contributions. In fact, because the CP violating phase is very small within the SM $\phi_s^{SM}\sim0.004$ and would be dominated by the NP contributions, the relation $\phi_s^{SM+NP}\simeq\phi^{J/\psi\phi}_s=-0.77^{+0.29}_{-0.37}\cup-2.36^{+0.37}_{-0.29}$ would be a good approximation within NP models~\cite{Lenz}. From Eq.~(\ref{assla}), one may find that any of the various NP models without providing significant modification to $\Gamma_{12}^s$ would be impossible to give a large negative $a^s_{\rm sl}$ as $-12.7\times10^{-3}$. While, one may note that there are large uncertainties in the experimental measurements for $a^s_{\rm sl}$ and $\dg_s$, which will hinder a definite conclusion.

Due to the relation $\phi_s^{SM+NP}\simeq\phi^{J/\psi\phi}_s$, which leads to $\cos\phi_s^{SM+NP}<\cos\phi_s^{SM}\simeq\cos(0.004)$, our prediction $\dg_s=(5.9\pm4.7)\times10^{-2} {\rm ps}^{-1}$ in S1 listed in Table~\ref{tabpred} is smaller than the SM result $(9.5\pm4.0)\times10^{-2} {\rm ps}^{-1}$ and the averaged experimental data $(15.4^{+5.4}_{-7.0})\times10^{-2} {\rm ps}^{-1}$~\cite{HFAG}. Taking into account of the large experimental error and theoretical uncertainties, they are still in good agreement. Recently, CDF and D0 Collaborations have updated their measurement
\ba
\dg_s&=&0.075\pm0.035\pm0.010\,{\rm ps}^{-1}\,,\quad {\rm CDF}~\cite{CDFnew}\\
\dg_s&=&0.15\pm0.06\pm0.01\,{\rm ps}^{-1}\,.\quad {\rm D0}~\cite{D0new}
\ea
One may find that the CDF result is much smaller than the D0 measurement, but close to our result $(5.9\pm4.7)\times10^{-2} {\rm ps}^{-1}$. The combination of CDF and D0 measurements is not available until now, which will be very important for further constraining the parameter space. Additionally, since there are two possible intervals for $\phi^{J/\psi\phi}_s$: $0>-0.77^{+0.29}_{-0.37}$~(S1)$>-\pi/2$ and $-\pi/2>-2.36^{+0.37}_{-0.29}$~(S2)$>-\pi$, our prediction $\dg_s>0$ in S1 and $\dg_s<0$ in S2 are listed in Table~\ref{tabpred}, which agree with the experimental results $15.4^{+5.4}_{-7.0}$ or $-15.4^{+7.0}_{-5.4}$, respectively.

\subsubsection*{Case~B: only color-octet operator}

In this case,  we consider the NP contributions induced by color-octet tensor operator solely. Our fitting results for the tensor parameters $\phi_{T8}$ and $|g_{T8}|$ are listed in the fourth and the fifth columns of Table~\ref{NPpara} and shown in Fig.~\ref{NPrange}~(b). Due to the different sign of ${\cal A}_{T8}^{RR}$ and ${\cal A}_{T1}^{RR}$, which could be found in Eqs.~(\ref{AT1}) and (\ref{AT8}), a negative $\phi_{T8}=-62.1^{\circ}\pm6.9^{\circ}$~($-74.2^{\circ}\pm7.1^{\circ}$) in S1~(S2) is demanded by the constraints from $\phi^{J/\psi\phi}_s$ and $a_{sl}^s$, which also can be seen from Figs.~\ref{obsnp}~($c$) and ($d$). Meanwhile, the strength of the color-octet tensor couplings $|g_{T8}|$ $\sim1.9(2.4)\times10^{-2}$ is much smaller than the color-singlet one $|g_{T1}|$ $\sim6.7(8.5)\times10^{-2}$ in S1~(S2). The situation of our numerical results and the effects of color-octet tensor operator for the observables in this case is similar to the ones of case~A.

In summary, we have studied the recently observed anomalies in $B_s^0-\bar{B}_s^0$ mixing with a set of possible tensor operators in a model-independent way. It is found that both $\phi^{J/\psi\phi}_s$ and $a^s_{\rm sl}$ could be bridged to the experimental data by the contributions of color-singlet or color-octet tensor operators with new weak phase $\phi_{T1}\sim28.0^{\circ}$~($15.6^{\circ}$) or $\phi_{T8}\sim-62.1^{\circ}$~($-74.2^{\circ}$) for the two possible solutions for the CP-violating phase $\phi^{J/\psi\phi}_s$  given by CDF and D0 measurements. So far, it is unknown that these tensor operators with coupling and weak phase determined in this letter could be generated in which realistic NP models. It surely deserves further studies. It is noted that the error bars in the measurement of  $\dg_s$, $\phi^{J/\psi\phi}_s$ and $a_{sl}^s$ are very large, which hinder us from deciphering color-singlet or color-octet tensor operator responsible for the anomalies. With the running LHC-b experiment, the refined measurements of these observables are expected to confirm or refute the NP effects.

\section*{Acknowledgments}
The work is supported by the National Science Foundation under contract
No.11075059 and the Startup Foundation for Doctors of Henan Normal University under contract
No.1006.

\end{document}